\documentclass[english,prl,groupedaddress,twocolumn,letterpaper,floatfix,showpacs]{revtex4}

\usepackage[T1]{fontenc}
\usepackage[latin9]{inputenc}
\usepackage{graphicx}
\usepackage{setspace}
\usepackage{amssymb}
\usepackage{amsmath}
\usepackage{subfig}

\newtheorem{theorem}{Theorem}[section]

\providecommand{\boldsymbol}[1]{\mbox{\boldmath $#1$}}

\newcommand{\ket}[1]{|{#1}\rangle}

\newcommand{\be}{\begin{equation}}
\newcommand{\ee}{\end{equation}}
\newcommand{\bea}{\begin{eqnarray}}
\newcommand{\eea}{\end{eqnarray}}

\usepackage{babel}

\begin{document}

\title{Path-Phase Information Complementarity for Interfering Particles through State-Discrimination}

\pacs{03.67.-a, 42.50.Dv, 42.50.Ex}

\author{Noam Erez}
\email{nerez@weizmann.ac.il}
\author{Daniel Jacobs}
\author{Gershon Kurizki}
\affiliation{Department of Chemical Physics,
Weizmann Institute of Science, Rehovot, 76100, Israel} 
\begin{abstract}
We analyze the trade-off between the amounts of information obtainable
on complementary properties of a qubit state by simultaneous measurements.
We consider a ``state discrimination'' scenario wherein the same measurements are repeated, but the
input states must be guessed in every run. We find a general complementarity
relation for path-phase guesses by any generalized measurements in this scenario. 
The counterpart of this input-output mutual information (MI) reveals a hitherto unknown aspect of complementarity.
\end{abstract}

\maketitle

\begin{figure}[t]
	\centering
	\includegraphics[width=\columnwidth]{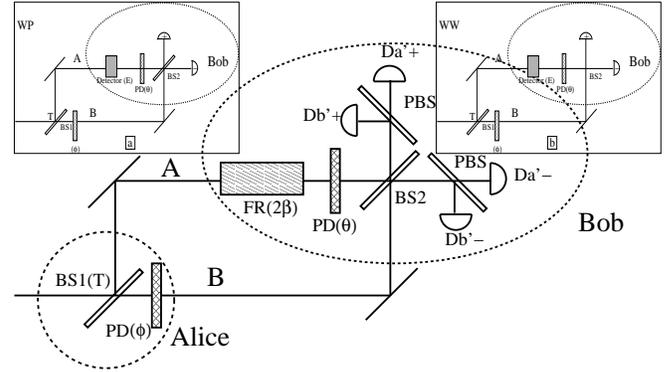}
  \caption{Setup for predictive complementarity game. 
The initial state is specified by a phase delay $\phi$ and transmissivity $T$ at BS1. Bob can either measure the phase 
 (inset a), or the path (inset b). Faraday rotator (FR) serves to correlate the path with the polarization.}
   \label{fig:three_beamsplits}
\end{figure}

As is well known, the measurement of one observable ``disturbs'' a
complementary observable, i.e., introduces uncertainty in it. 
A complementarity or duality relation has been derived\cite{refs1, EB}
and experimentally verified\cite{ref_exp}
for Hilbert space of dimensionality 2. This relation quantifies
 path predictability versus fringe visibility of a particle in a Mach-Zehnder
interferometer (MZI) with a partly efficient which-path detector
(Fig. \ref{fig:three_beamsplits}a ).
This relation reads:
\begin{equation}
\label{eq:std_comp}
D^{2}+V^{2}\leq1
\end{equation}
The path distinguishability, $D$, is related to the which-way probability,
$\mathcal{P}_{\rm ww}$, of guessing the path correctly for a {\em known} input state and
a which-way detector of efficiency (reliability) $E\leq1$, while the fringe visibility, $V$, is related to the which-phase probability, $\mathfrak{\mathcal{P}}_{\rm WP}$, of guessing correctly which MZI port the
particle will exit through (for an optimal choice of the phase between the arms) \cite{EB, TIE} 
\begin{equation}
\mathfrak{\mathcal{P}}_{\rm WW}=\frac{1+D}{2},~~\mathfrak{\mathcal{P}}_{\rm WP}=\frac{1+V}{2}.
\end{equation}
Yet, in this setup, the WW and WP probabilities refer to two {\em alternative}
measurements\cite{Luis}. Indeed, $\mathcal{P}_{\rm WP}$ is the probability of
predicting correctly where the particle will exit (inset (a) of Fig. \ref{fig:three_beamsplits}a ).
By contrast, $\mathfrak{\mathcal{P}}_{\rm WW}$ is
operationally meaningful only in a measurement (inset (b) of the figure). 
\ref{fig:three_beamsplits}) where the exit beam splitter of the MZI is removed,
because only then can the readout of the partly efficient WW detector be
verified.  Thus, in the scheme of Fig. \ref{fig:three_beamsplits}, 
simultaneous guesses of path and phase cannot be verified or falsified in the
same predictive experiment, i.e., either $\mathcal{P}_{\rm WP}$ or
$\mathcal{P}_{\rm WW}$ must represent a counterfactual probability.
We may think of this ``predictive'' duality as a constraint on the optimal strategies in a single-player game, in which the player (Bob) knows the initial state and the experimental setup and tries to guess the outcome of each measurement.

\begin{figure}[t]
    \subfloat[]{
        \label{fig:rect_states}
        \includegraphics[width=.35\columnwidth]{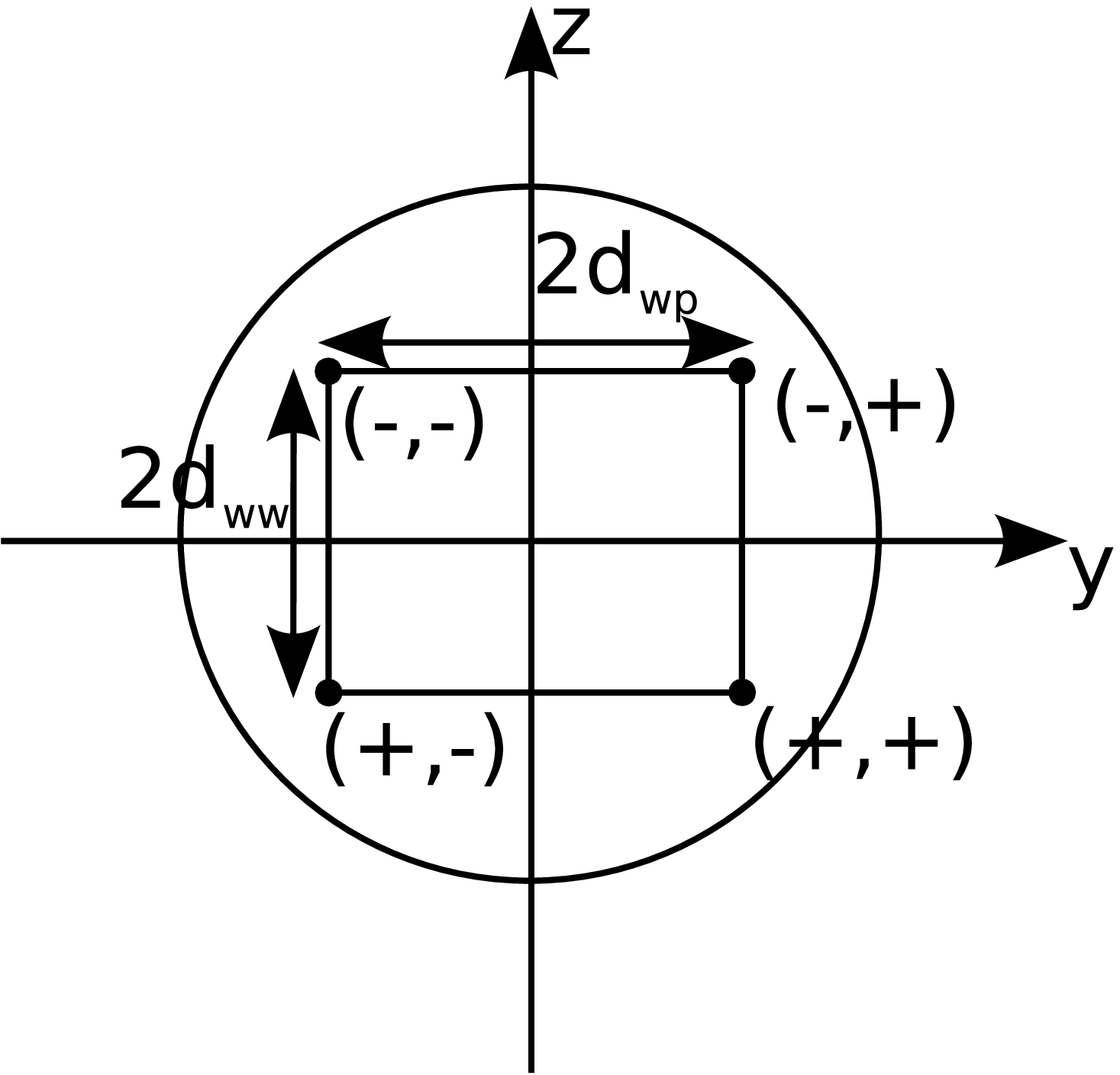}}
    \subfloat[]{
        \label{fig:8_states}
        \includegraphics[width=.55\columnwidth]{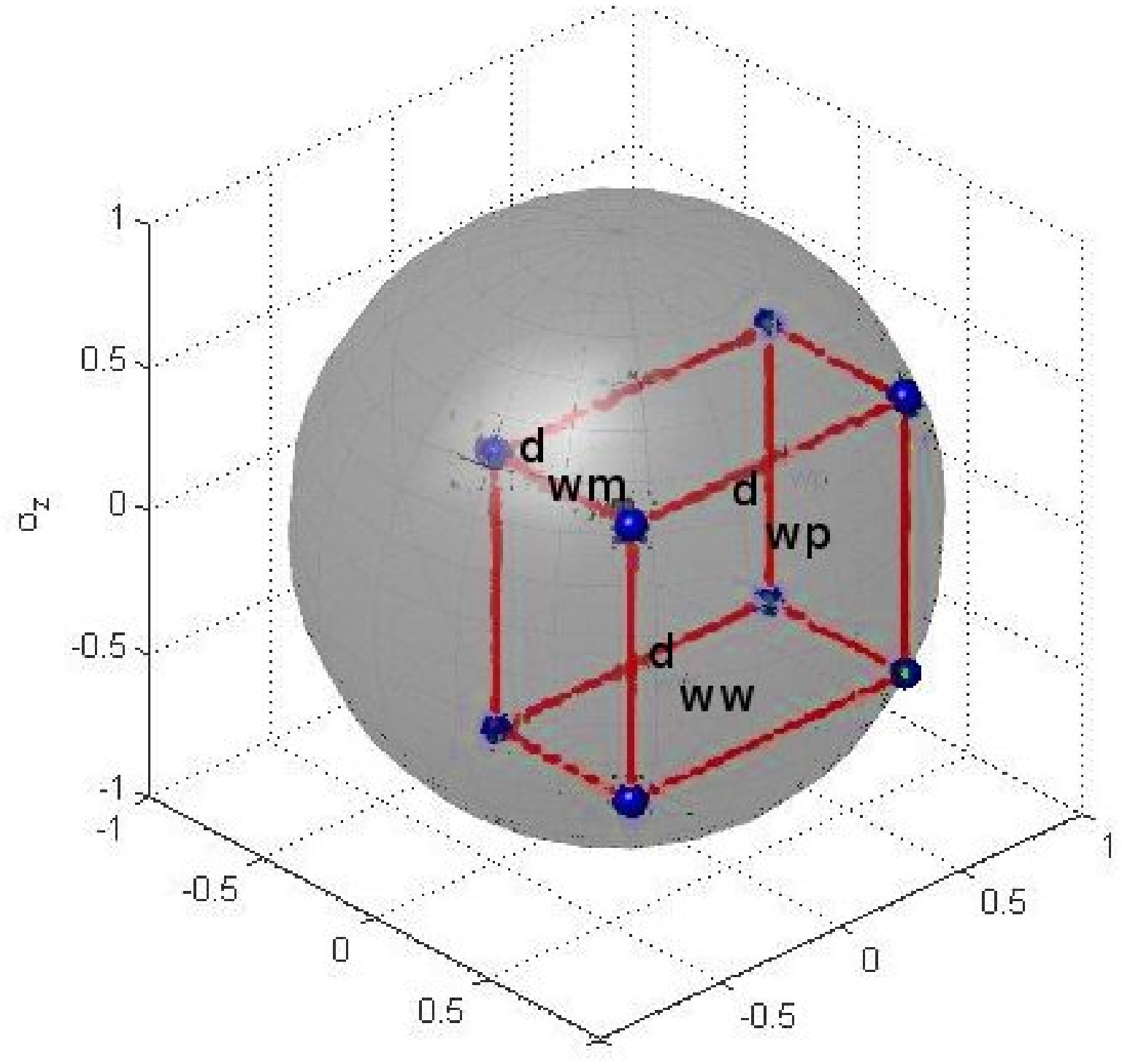}}
    \hspace{0.1\linewidth}
    \subfloat[]{
        \label{fig:beamsplit_tunable}
        \includegraphics[width=.45\columnwidth]{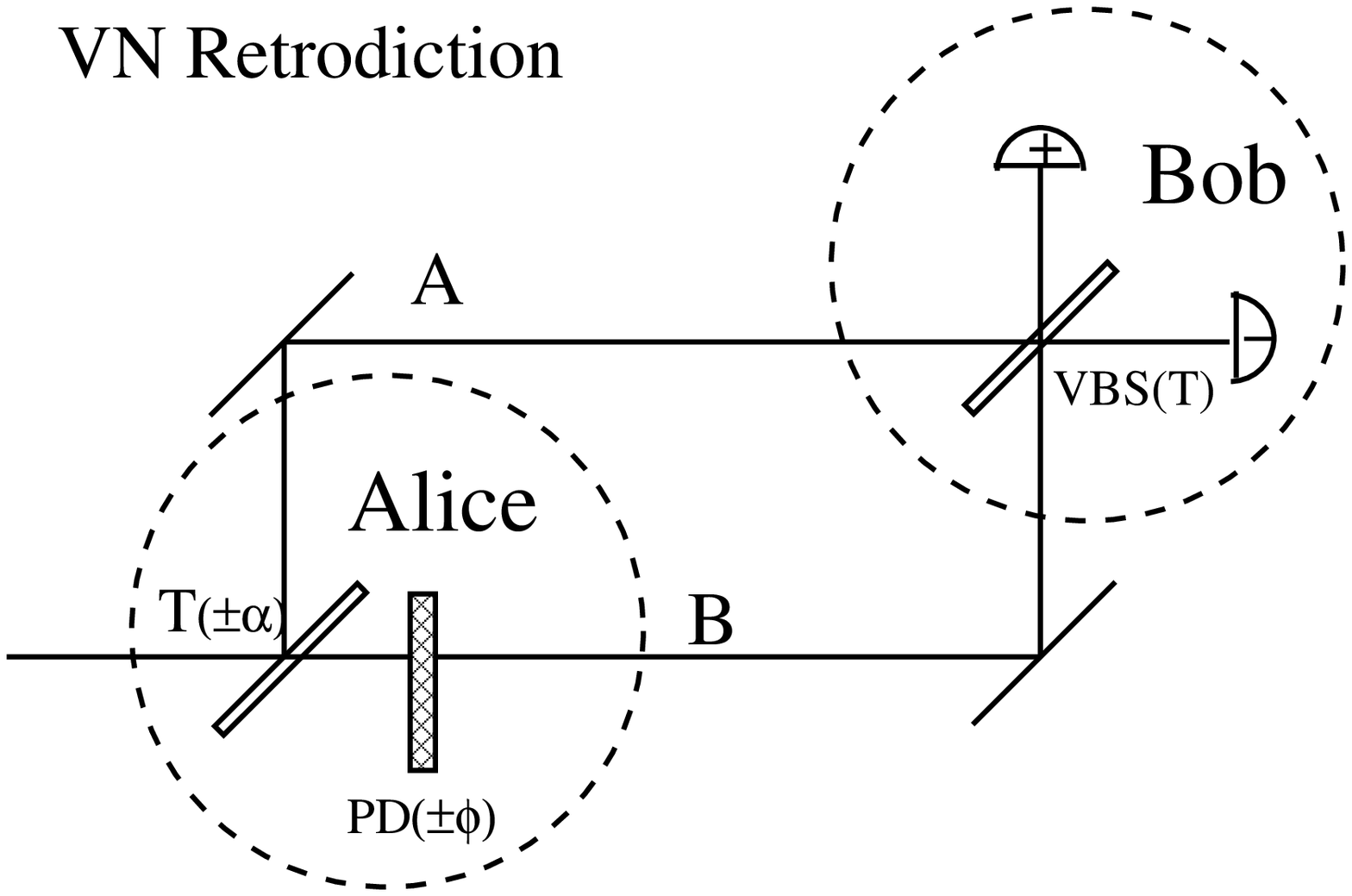}}
    \subfloat[]{
        \label{fig:multiple_bs}
        \includegraphics[width=.45\columnwidth]{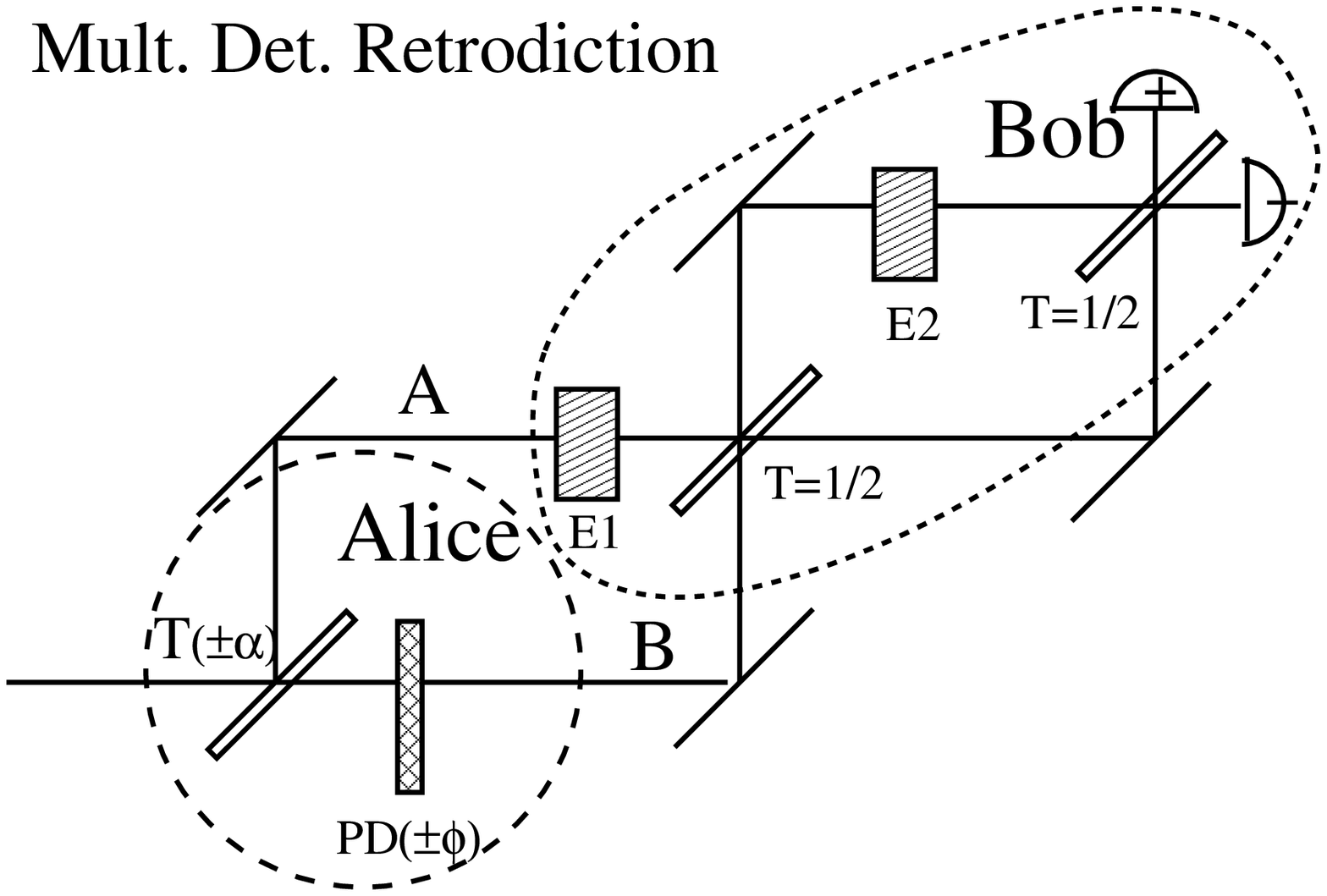}}
    \caption{
    (a) Bloch representation of Alice's four alternative WW-WP input states (labeled by  $b_{\rm ww}$ and $b_{\rm wp}$), with distances $d_{\rm ww}$ and $d_{\rm wp}$. 
    (b) Bloch representation of 8 WW-WP-WM input states. 
    (c) VN scheme (tunable output BS). 
    (d) Multiple inefficient detector scheme for mixed state discrimination.}
    \label{fig:three_impls}
\end{figure}

Is it possible to obtain a duality relation for path and phase information such that 
both have {\em simultaneous} operational meaning in each experimental run? As we show,
such a relation can indeed be given in the context of quantum {\em state discrimination},
namely, measurements aimed at optimally guessing the initial state out of a set of possible
states\cite{HB}.
In contrast to the ``predictive'' WW-WP duality, the proposed ``retrodictive'' duality described below is a bound on the optimal strategies in a two-player game: the guessing by Bob which of the several alternative input states had been prepared by Alice prior to the one measurement Bob performed. 
The precise rules of the game for this state-discrimination scenario are as follows:

(1) Alice randomly chooses to prepare the particle in one of the four input states (Fig. \ref{fig:rect_states}):
\begin{equation}
\label{eq:inputs}
\ket{b_{\rm ww},b_{\rm wp}}_{\alpha,\phi}\equiv T(b_{\rm ww}\alpha)\ket{A}+e^{b_{\rm wp}i\phi}T(-b_{\rm ww}\alpha)\ket{B}
\end{equation}
Here $\ket{A,B}$ are the path states (represented by qubit states
$\ket{\sigma_{z}=\pm1}$), $T(\pm{}b_{\rm ww}\alpha)=\cos\left(\frac{\pi}{4}\pm\frac{b_{\rm ww}\alpha}{2}\right)$
the corresponding amplitudes, and  $e^{b_{\rm wp}i\phi}$ their
relative phase factor. The four input states correspond to the choices of the parameters $\left(b_{\rm ww}=\pm1,b_{\rm wp}=\pm1\right)$.

(2) Bob receives the qubit, and after performing a measurement of his choice, tries
to guess the values of the two bits $b_{\rm ww},b_{\rm wp}$ (which are statistically
independent), i.e., guess which of the four possible input states was chosen by
Alice. 

For a given choice of Alice's parameters $(\alpha, \phi)$, each strategy that Bob adopts yields probabilities $P_{\rm WW}$ and $P_{\rm WP}$ to correctly guess $b_{\rm ww}$ and $b_{\rm wp}$, respectively (Roman font is henceforth used for $P_{\rm WW}$ and $P_{\rm WP}$, as opposed to the calligraphic font for ``predictive'' probabilities $\mathcal{P}_{\rm ww}$ and $\mathcal{P}_{\rm wp}$ above). Bob's strategy is ``Pareto optimal''\cite{Pareto}
if there is no other strategy that yields a pair of probabilities $(P_{\rm WW}$, $P_{\rm WP})$ such that one is strictly better and the other not worse than its counterpart. 
The set of all optimal pairs is called the ``Pareto Frontier''. 

The optimal probability of distinguishing two states of a qubit with density operators $\rho_1$ and $\rho_2$ is given by $1/2$ plus their trace distance\cite{nielsen521qca}: $d_{\rm Trace}(\rho_1,\rho_2)\equiv\frac{1}{2}{\rm Tr}|\rho_1-\rho_2|$. This is half the Euclidean distance between
the corresponding Bloch vectors. Let us define $d_{\rm ww}\equiv
d_{\rm Trace}(|b_{\rm ww}=+1,b_{\rm wp}\rangle,|b_{\rm ww}=-1,b_{\rm wp}\rangle),d_{\rm wp}\equiv
d_{\rm Trace}(|b_{\rm ww},b_{\rm wp}=+1\rangle,|b_{\rm ww},b_{\rm wp}=-1\rangle)$.
Here $d_{\rm ww}$ denotes the which-way distinguishability of the
set of inputs, while $d_{\rm wp}$ is the which-phase distinguishability.

The Bloch vector corresponding to the input state
$\mathbf{b}^{\rm in}_i=$ $|b_{\rm ww},b_{\rm wp}\rangle_{\alpha,\phi}$ in Eq. (\ref{eq:inputs}) is:
\begin{equation}
\label{eq:rb1b2}
\mathbf{r}_i=d_{0}\mathbf{\hat{x}}+b_{\rm wp}d_{\rm wp}\mathbf{\hat{y}}+b_{\rm ww}d_{\rm ww}\mathbf{\hat{z}},
\end{equation}
where $\mathrm{\mathit{d_{0}=\cos\alpha\cos\phi;\, d_{\rm ww}=\sin\alpha;\, d_{\rm wp}=\cos\alpha\sin\phi.}}$
The set of Alice's allowed input states form a rectangle on the Bloch Sphere with dimensions given by $2d_{\rm ww}$ and $2d_{\rm wp}$. (Fig. \ref{fig:rect_states}). 

Bob is allowed to perform generalized measurements (POVMs), as well as projective (Von Neumann) ones, by letting the particle interact with an ancillary system (ancilla) and then performing a projective
measurement on both together (the WW detector in Fig. \ref{fig:three_beamsplits} is just such an ancilla).
Any POVM can be represented by a set of operators, $\left\{ A_{i}\right\} _{i=1...N}$,  satisfying\cite{peres1995qtc} $A_{i}\geq0,A_{i}^{\dagger}=A_{i},\sum_i A_{i}=1$. Operationally, this means that when performing the corresponding generalized measurement on a system initially in state $\rho$, the $i$th ``measurement'' outcome appears with probability ${\rm Tr}\left\{A_i \rho\right\}$.

\textbf{Lemma} A POVM $\left\{ A_{j}\right\}$ on a qubit can be described as a collection of weighted points in the Bloch Ball: $A_{j}=\mu_{j}\frac{1+\mathbf{R}_{j}\cdot\mathbf{\boldsymbol{\sigma}}}{2}$, with
$\mu_{j}\geq0,\left\Vert \mathbf{R}_{j}\right\Vert \leq1$. Furthermore, every such POVM has a ``refinement'' 
consisting of weighted points on the Bloch {\em Sphere} ($\left\Vert \mathbf{R}_{j}\right\Vert =1$), such that all information on the original POVM can be retrieved from it. Thus, without loss of generality, we shall assume a representation of the latter form. 
The proof of this lemma is given in the Supplement.

By assumption, all inputs $\mathbf{b}^{in}_{\mathbf{i}}$ are equally probable: $p_{\mathbf{i}}=\frac{1}{4}$.
The joint input-output distribution is then: $p_{\mathbf{i},j}=p_{\mathbf{i}}Tr\left\{ \rho_{\mathbf{i}}A_{j}\right\}=$ $\frac{\mu_{j}}{8}(1+d_{0}x_{j}+i_{\rm wp}y_{j}d_{\rm wp}+i_{\rm ww}z_{j}d_{\rm ww})$, where
$\mathbf{R}_{j}= \left( x_j,y_j,z_j \right)$. 
\begin{theorem}
\label{th:ellipse} For any POVM, the WW and WP probabilities satisfy:
\begin{equation}
\label{eq:ellipse}
\left(\frac{2P_{\rm WW}-1}{d_{\rm ww}}\right)^{2}+\left(\frac{2P_{\rm WP}-1}{d_{\rm wp}}\right)^{2} \leq 1
\end{equation}
Equality holds iff all the Bloch vectors have the form $\mathbf{R}_j = (0, \pm \sqrt{1-z_0^2}, \pm z_0)$, with $z_0 \in [0,1]$, and the corresponding weights $\mu_{\pm,\pm}$ satisfy: 
$\mu_{++}=\mu_{--}=$ $\mu,~~\mu_{+-}=$ $\mu_{-+}=1-\mu$ for some $\mu \in [0,1]$.
\end{theorem}

\textbf{Outline of proof}
As shown in detail in the Supplement, the probability of correctly guessing $b_{\rm ww}$ is given by:
\be
P_{\rm WW}=\sum_{j}\max\left\{ p_{i_{\rm ww},j},p_{-i_{\rm ww},j}\right\}= \frac{1}{2}\left(1+\sum\frac{\mu_{j}}{2}\left|z_{j}\right|d_{\rm ww}\right), 
\label{eq:PWWb}
\ee
\begin{equation}
P_{\rm WP}=\frac{1}{2}\left(1+\sum\frac{\mu_{j}}{2}\left|y_{j}\right|d_{\rm wp}\right).
\label{eq:13b}
\end{equation}
Combining these two equations, we have
\begin{equation}
\left(\frac{2P_{\rm WW}-1}{d_{\rm ww}}\right)^{2}+\left(\frac{2P_{\rm WP}-1}{d_{\rm wp}}\right)^{2} \leq \sum\frac{\mu_{j}}{2}\left(z_{j}^{2}+y_{j}^{2}\right) \leq 1.
\end{equation}
$\blacksquare$

In what follows we shall restrict ourselves to measurements (POVMs) which are Pareto optimal, unless stated otherwise.
Two particular instances of this optimal class are of special interest.
The case $\mu \in \{0,1\}$ describes a Von-Neumann measurement in the $x-z$ plane (this can be realized by
choosing the phase delays of the output beam splitter of the MZI appropriately).
The case $\mu = \frac{1}{2}$ describes the measurement with a WW detector, as in Fig. \ref{fig:three_beamsplits}a.
 
For the WW-detector assisted measurement (Fig. \ref{fig:three_beamsplits}a )
, the probabilities are determined by the detector efficiency, $E$: 
$z_0=E$. 
Likewise, in the Von-Neumann measurement corresponding to an output BS with transmissivity $\frac{1}{2}\leq T_{\rm Out} \leq 1$ and correctly chosen input phases (Fig. \ref{fig:beamsplit_tunable}), $z_0 = 1-2T_{\rm Out}$.

The (unique) measurement minimizing the overall probability of error (in guessing
both path and phase) is also Pareto optimal:
\begin{theorem}
\label{th:prob_err}
A POVM that maximizes the probability of guessing the path and phase simultaneously is necessarily WW-WP
Pareto-optimal.
\end{theorem}
\textbf{Proof:} A similar calculation to that used in the proof of
Theorem \ref{th:ellipse} gives the probability of correctly guessing the input
state, $P_{c}=\frac{1}{2}\left(P_{\rm WW}+P_{\rm WP}-\frac{1}{2}\right)$. Clearly,
increasing one of the partial probabilities without reducing the other implies
improving the average $\blacksquare$


{\bf \em WW- and WP-Information complementarity. }
These results can be recast in information-theoretic terms. 
Let $\mathbf{b}^{\rm in}=(b_{\rm ww},b_{\rm wp})$ be the two random (statistically independent) bits Alice chooses for her \emph{input} state, as before, and denote the observables measured by Bob by $\mathbf{b^{\rm out}}$. 
This set of {\em classical} stochastic variables (albeit related by a quantum channel) allows the definition of the WW(WP)- output information as the mutual information between $b_{\rm ww(wp)}$ and $b^{out}$: $I_{WW} \equiv I(b_{\rm ww}: \mathbf{b^{\rm out}});~~I_{WP} \equiv I(b_{\rm wp}: \mathbf{b^{\rm out}}).$

To separate out the WW correlation explicitly, we introduce a new stochastic variable
constructed out of the fundamental ones:
\be
WW = \left\{ 
\begin{array}{ll}
			1 & P(b_{\rm ww},\mathbf{b}^{\rm out}) > P(b_{\rm ww},\mathbf{b}^{\rm out}) \\
			0  & {\rm otherwise} 
\label{eq:9}
\end{array} 
\right.
\ee
where $P(b_{\rm ww},\mathbf{b}^{\rm out})$ is the apriori probability for 
$b_{\rm ww},\mathbf{b}^{\rm out}$ to take the values which actually occurred. 
We see from Eq. \eqref{eq:9} that $P_{\rm WW} = P(WW = 1).$
This also implies $H(WW)=H_2(P_{\rm WW})$, where $H_2(x)=H(x,1-x)$ is the binary entropy.
We define an analogous variable $WP$, such that $P_{WP} = P(WP = 1)$.

Now, given the value of $\mathbf{b}^{\rm out}$, $b_{\rm ww}$ and $WW$ determine each other, 
so that they are interchangeable, in the joint entropy 
$H(b_{\rm ww},\mathbf{b}^{\rm out})=H(WW,\mathbf{b}^{\rm out})$. Since for optimal measurements,
$WW$ is stochastically independent of $\mathbf{b}^{\rm out}$, as can be seen from $P_{\rm WW}=\frac{1}{2}\left(1+z_0 d_{\rm ww}\right)$ (see Supplement), it follows that 
$H(b_{\rm ww},\mathbf{b}^{\rm out})=H(WW,\mathbf{b}^{\rm out})=H(WW)+H(\mathbf{b}^{\rm out})$.
By definition, the mutual information $I(b_{\rm ww}: \mathbf{b^{\rm out}}) \equiv 
H(b_{\rm ww})+H(\mathbf{b^{\rm out}})-H(b_{\rm ww},\mathbf{b^{\rm out}})=H(b_{\rm ww})-H(WW)$. From this last equation
(and its WP analog) follow the intuitively appealing relations:
\be
I_{\rm WW} = 1 - H_2(P_{\rm WW}),~~I_{\rm WP} = 1 - H_2(P_{\rm WP}).
\label{eq:10}
\ee 

We note that $I_{\rm WW}$ and $I_{\rm WP}$ are monotonically increasing functions of
$P_{\rm ww}$ and $P_{\rm wp}$, respectively (because $P_{\rm ww}, P_{\rm wp} \in [\frac{1}{2},
1]$). Thus, the $P_{\rm WW}$-$P_{\rm WP}$ complementarity of Eq. (\ref{eq:ellipse}) implies a similar trade-off for
$I_{\rm WW}$ and $I_{\rm WP}$.
The amount of information about Alice's input settings contained in Bob's measurement results is given by the mutual information between them: $I_{\rm in-out} = I(\mathbf{b}^{\rm in}:\mathbf{b}^{\rm out}).$
The relation between $I_{\rm in-out}$ and $I_{\rm WW},~I_{\rm WP}$ is given by the following:
\begin{theorem}
\label{th:Idecomp} 
\be
I_{\rm in-out} = I_{\rm WW}+I_{\rm WP}+I_{\rm WW:WP}.
\label{eq:26}
\ee
\end{theorem}
\textbf{Proof: }(1) For the important special case $\mu = \frac{1}{2}$ (WW-detector scheme), 
the output observables also consist of two bits 
$\mathbf{b^{\rm out}}=(b_{\rm ww}^{out},b_{\rm wp}^{out})$, which turn out to be statistically independent. As the notation suggests, the WW output bit (obtained from the reading of the WW detector) has non-zero mutual information with $b_{\rm ww}$, but not with $b_{\rm wp}$, and conversely for the WP output bit.
From
\bea
\label{eq:27}
& &I(b_{\rm ww},b^{\rm out}_{\rm ww}:b_{\rm wp},b^{\rm out}_{\rm wp}) \equiv \nonumber \\
& &H(b_{\rm ww},b^{\rm out}_{\rm ww})+H(b_{\rm wp},b^{\rm out}_{\rm wp}) -H(b_{\rm ww},b_{\rm wp};b^{\rm out}_{\rm ww},b^{\rm out}_{\rm wp}),
\eea
and 
\bea
& &H(b_{\rm ww},b^{\rm out}_{\rm ww})=H(WW)+1,H(b_{\rm wp},b^{\rm out}_{\rm wp})=H(WP)+1 \nonumber \\
& &H(b_{\rm ww},b_{\rm wp};b^{\rm out}_{\rm ww},b^{\rm out}_{\rm wp}) = H(WW,WP)+2,
\eea
we obtain:
\bea
& &I(b_{\rm ww},b^{\rm out}_{\rm ww}:b_{\rm wp},b^{\rm out}_{\rm wp}) \equiv \nonumber \\
& &H(WW)+H(WP) -H(WW,WP) = I(WW:WP).
\eea
Comparing
\bea
& &I_{\rm in-out}=I(b_{\rm ww},b_{\rm wp}:b^{\rm out}_{\rm ww},b^{\rm out}_{\rm wp}) \equiv \nonumber \\
& &H(b_{\rm ww},b_{\rm wp})+H(b^{\rm out}_{\rm ww},b^{\rm out}_{\rm wp}) -H(b_{\rm ww},b_{\rm wp};b^{\rm out}_{\rm ww},b^{\rm out}_{\rm wp}),
\label{eq:15}
\eea
with Eq. \eqref{eq:27}, 
and using the independence of $b_{\rm ww},b_{\rm wp}$ and of $b^{\rm out}_{\rm ww},b^{\rm out}_{\rm wp}$ to decompose:
$H(b_{\rm ww},b_{\rm wp})=H(b_{\rm ww})+H(b_{\rm wp}),~H(b^{\rm out}_{\rm ww},b^{\rm out}_{\rm wp})=H(b^{\rm out}_{\rm ww})+H(b^{\rm out}_{\rm wp})$,
we get Eq. \eqref{eq:26}. 

The third term, $I_{WW:WP}$, is a novel corollary of our treatment. This ``cross-information''has its origin in the fact
that, although $b_{\rm ww}$ is independent of $b_{\rm wp}$ and of $b_{\rm wp}^{out}$, and
$b_{\rm ww}^{out}$ is also independent of $b_{\rm wp}$ and of $b_{\rm wp}^{out}$,
$b_{\rm ww}^{out}$ and $b_{\rm ww}$ together tell us something about $b_{\rm wp}^{out}$ and
$b_{\rm wp}$.  Namely, a correct (incorrect) WW guess implies lower (higher)
probability of correct WP guess. If Bob is allowed to bet on WW first, and 
is then told the outcome, this will not change the value of $b_{\rm wp}$ he bets on, but will affect
the odds of his guessing correctly! This is a {\em hitherto unnoticed subtle form of complementarity}.


(2) In the general case ($\mu \in [0,1]$), $\mathbf{b}^{out}$ does not decompose neatly into WW and WP parts, and proceeding as above, one ends up with the relation:
$I_{\rm in-out} = $ $I_{\rm WW} + I_{\rm WW} +$ $\left[I\left(b_{\rm ww},\mathbf{b}^{\rm out}:b_{\rm wp},\mathbf{b}^{\rm out}\right) - H\left(\mathbf{b}^{\rm out}\right) \right]$. 
The expression in the square brackets is known as the conditional mutual information:
$I\left(b_{\rm ww}:b_{\rm wp}|\mathbf{b}^{\rm out}\right)$ and is equal to $I_{\rm WW:WP}$, as required.
$\blacksquare$


\begin{figure}[htp]
    \subfloat[]{
        \label{fig:i_tradeoff}
        \includegraphics[width=.4\columnwidth]{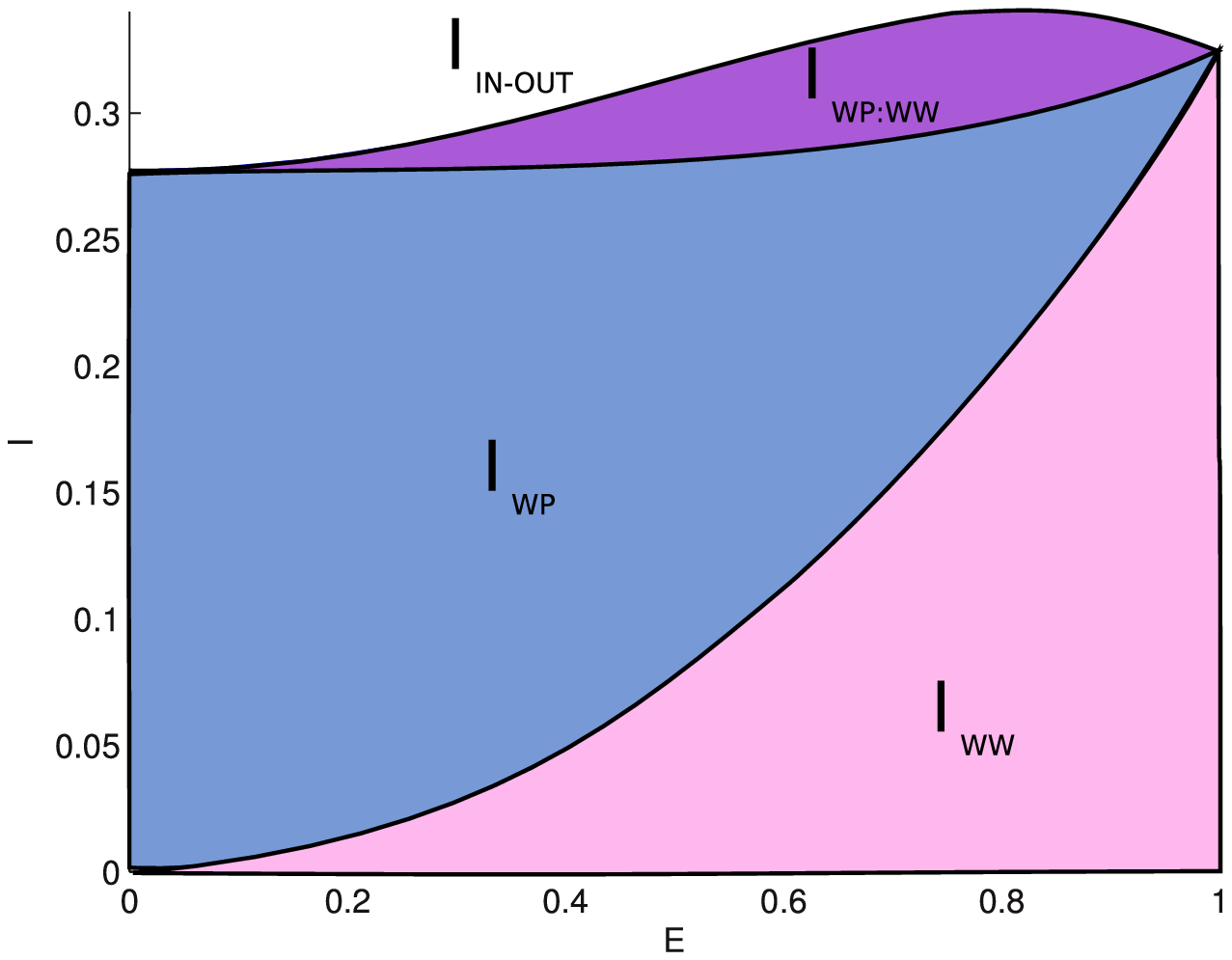}}
		\subfloat[]{ 
        \label{fig:color_3ps}
        \includegraphics[width=.4\columnwidth]{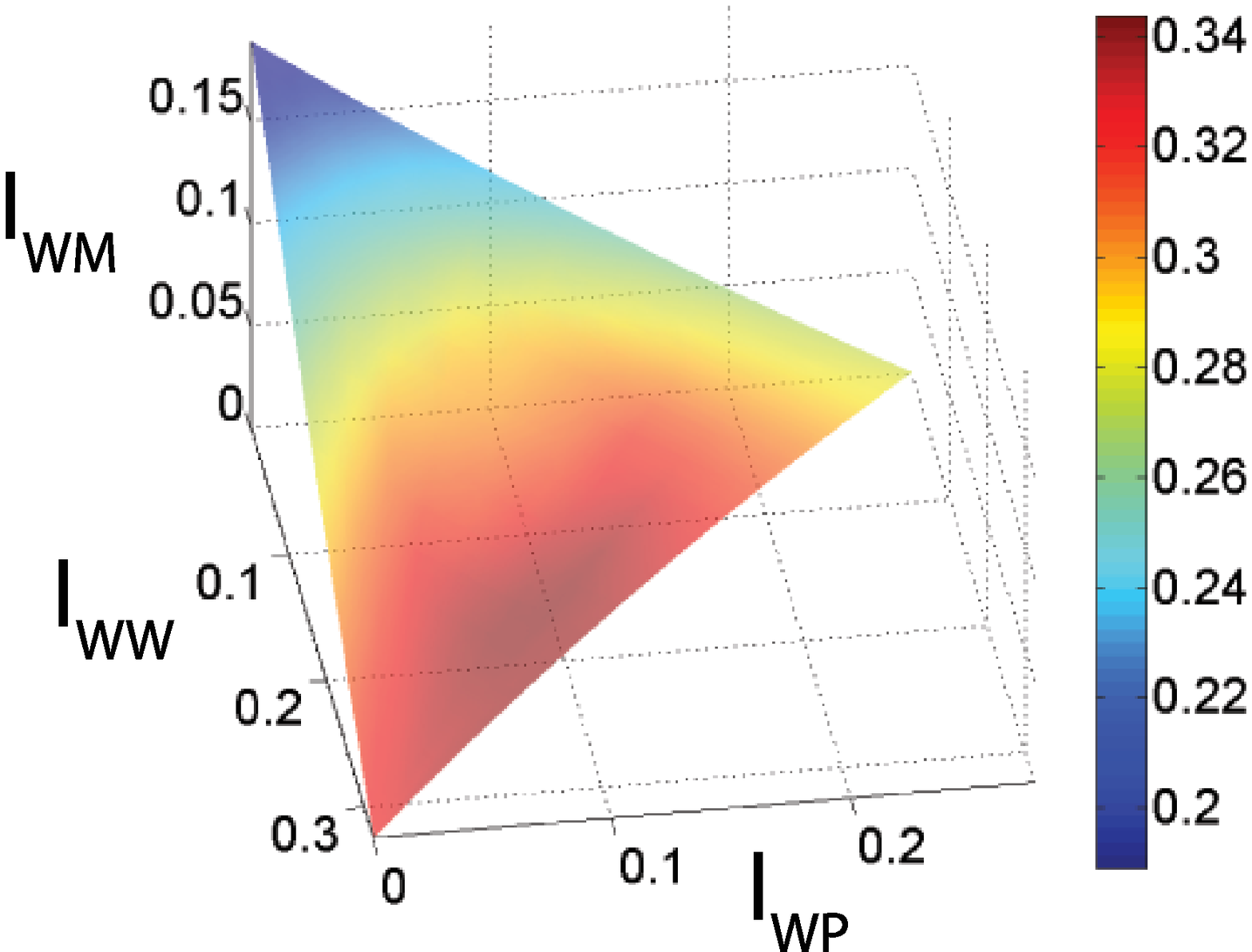}}
    \caption{
    (a) Illustration of \eqref{eq:26} parametrized by the detector efficiency, $E$ in a WW-detector scheme (input parameters: $d_{\rm ww}=.65, d_{\rm wp}=.6$). 
    (b) Pareto optimal surface  with color representing the magnitude of $I_{in-out}$ (input parameters: $d_{\rm ww}=.65$, $d_{\rm wp}=.6$, $d_{\rm wm}=0.5$).}
\label{fig:3}
\end{figure}


{\bf \em Complementarity for mixed states}
The geometry of the trace-distance suggests the following generalization of the
problem to 3 independent input parameters for mixed states. We now consider 8
input states $\rho_{b_{\rm ww}=\pm{1},b_{\rm wp}=\pm{1},b_{\rm wm}=\pm{1}}$ corresponding
to the Bloch vectors:
\begin{equation}
\mathbf{r}_{b_{\rm ww},b_{\rm wp},b_{\rm wm}}=d_{0}\mathbf{\hat{x}}+b_{\rm wm}d_{\rm wm}\hat{\mathbf{x}}+b_{\rm wp}d_{\rm wp}\mathbf{\hat{y}}+b_{\rm ww}d_{\rm ww}\mathbf{\hat{z}},
\end{equation}
where $d_{\rm wm}$ is the mixing distance (a measure of impurity), defined
similarly to $d_{\rm ww}$ and $d_{\rm wp}$, and the distances satisfy:
$d_{\rm ww}^{2}+d_{\rm wp}^{2}+d_{\rm wm}^{2}\leq1$ and
$d_{0}=\sqrt{1-\left(d_{\rm ww}^{2}+d_{\rm wp}^{2}\right)}-d_{\rm wm}$. The set of all
eight states is now a rectangular box within the Bloch Ball (Fig. \ref{fig:8_states}).
The Pareto frontier is now an ellipsoid:
\begin{equation}
\label{eq:ellipsoid}
\left(\frac{2P_{\rm WW}-1}{d_{\rm ww}}\right)^{2}+\left(\frac{2P_{\rm WP}-1}{d_{\rm wp}}\right)^{2}+\left(\frac{2P_{\rm WM}-1}{d_{\rm wm}}\right)^{2}=1.
\end{equation}
Schemes for path-phase-``mixedness'' complementarity are shown in Fig. \ref{fig:beamsplit_tunable} and \ref{fig:multiple_bs} (see supplement). 
The input-output mutual information for optimal schemes is:
\begin{equation}
\label{eq:3_way_info_comp}
I_{\rm in-out}= I_{\rm WW}+I_{\rm WP}+I_{\rm WM}+ C_{\rm WW:WP:WM}.
\end{equation}
Here $C_{\rm WW:WP:WM}$ is the ``total correlation'' $C(WW:WP:WM)$ defined by:
$C(X,Y,Z)\equiv H(X)+H(Y)+H(Z)-H(X,Y,Z).$ It is the mixed-state counterpart of the novel cross-information
term $I_{\rm WW:WP}$ in \eqref{eq:26} (see discussion following \eqref{eq:15}).


To conclude, while the ``predictive'' distinguishability-visibility duality holds for \emph{two}
alternative measurements (itself a manifestation of complementarity), our state-discrimination scenario 
allows for only one type of measurement: we have more preparations, instead. This
scenario yields the generalized complementarity and mutual information (MI)
relations (\ref{eq:ellipse}) and \eqref{eq:26}, and their mixed input
generalization, (\ref{eq:ellipsoid}) and (\ref{eq:3_way_info_comp}).  These
results show that complementarity may be reformulated as an information
tradeoff obtained on complementary properties (path and phase) in a single
measurement.  The structure of this information tradeoff is richer than
previously thought, as it allows for ``cross information'': information gained on the odds of WW guess given
the WP result (or vice versa).
Since the goal of this work is to allow for {\em simultaneous} WW and WP MI, our scenario is {\em not} the 
sequence-reversed version of standard predictive duality. However, the latter scenario is of great interest
to quantum cryptography\cite{wu2009ci}. As shown in \cite{Erez:arXiv0903.1921}, one can derive Pareto-optima for the latter problem from those of the present one quite simply. Hence, the tight bound on MI in \eqref{eq:26}, \eqref{eq:3_way_info_comp} is expected to be useful for improving the corresponding one in the cryptographic setting.

We acknowledge the support of the EC, GIF and ISF.




\section*{Supplement}

\textbf{Lemma} A POVM $\left\{ A_{j}\right\}$ on a TLS, can be described as a collection of weighted points in the Bloch Ball: $A_{j}=\mu_{j}\frac{1+\mathbf{R}_{j}\cdot\mathbf{\boldsymbol{\sigma}}}{2}$, with
$\mu_{j}\geq0,\left\Vert \mathbf{R}_{j}\right\Vert \leq1$. Furthermore, every such POVM has a ``refinement'' 
consisting of weighted points on the Bloch {\em Sphere} ($\left\Vert \mathbf{R}_{j}\right\Vert =1$), such that all information on the original POVM can be retrieved from it.

\textbf{Proof:} Any hermitian operator $A$ on a qubit's Hilbert space is a
linear combination of the Pauli matrices and the identity:
$A=\frac{\mu}{2}1+\mathbf{R}\cdot\mathbf{\boldsymbol{\sigma}}$. If $A$ is
further required to be positive, then we must have $\left\Vert
\mathbf{R}\right\Vert \leq\frac{\mu}{2}$, as can be seen from the requirement
$\left\langle
\hat{\mathbf{R}}\cdot\sigma=-1\right|A\left|\hat{\mathbf{R}}\cdot\sigma=-1\right\rangle
\geq0$.  Hence
$A_{i}=\mu_{i}\frac{1+\mathbf{R}_{i}\cdot\mathbf{\boldsymbol{\sigma}}}{2}$ with
$\mu_{i}\geq0,\left\Vert \mathbf{R}_{i}\right\Vert \leq1$, which is, by
definition, in the Bloch ball. 
The condition that the operators sum to the identity implies
$\frac{1}{2}\sum\mu_{i}=1;\sum\mu_{i}\mathbf{r}_{i}=0$. Also notice that every
POVM has a ``refinement'' consisting of weighted points on the Bloch Sphere,
such that all information on the original POVM can be retrieved from its
refinement. The refinement is obtained by replacing each operator $A_{i}$ whose
Bloch vector lies inside the ball
($\left(\mu_{i},\mathbf{R}_{i}\right),|\mathbf{R}_{i}|<1$), by the pair
$\left\{ A_{i}^{\pm}\right\} $, where
$A_{i}^{\pm}\leftrightarrow\left(\mu_{i}\frac{1\pm|\mathbf{R}_{i}|}{2},\pm\hat{\mathbf{R}}_{i}\right).$

$\blacksquare$

\textbf{Theorem \ref{th:ellipse}}
For any POVM, the WW and WP probabilities satisfy:
\begin{equation}
\left(\frac{2P_{\rm WW}-1}{d_{\rm ww}}\right)^{2}+\left(\frac{2P_{\rm WP}-1}{d_{\rm wp}}\right)^{2} \leq 1
\label{eq:5}
\end{equation}
Equality holds iff the Bloch vectors all have the form $\mathbf{R}_j = (0, \pm \sqrt{1-z_0^2}, \pm z_0)$, with the same $z_0 \in [0,1]$, and the corresponding weights $\mu_{\pm,\pm}$ satisfy: 
$\mu_{++}=\mu_{--}=$ $\mu,~~\mu_{+-}=$ $\mu_{-+}=1-\mu$ for some $\mu \in [0,1]$.

\textbf{Proof}: 
Let $\mathbf{i}=1 \ldots 4$ stand for each possible $(i_{\rm ww},i_{\rm wp})$ pair, as described in Eq. (\ref{eq:inputs}). 
By assumption all inputs $\mathbf{i}$ are equally probable, and we use the notation: $p_{\mathbf{i}}=\frac{1}{4}$
For the joint input-output distribution: $P_{\mathbf{i},j}=p_{\mathbf{i}}Tr\left\{ \rho_{\mathbf{i}}A_{j}\right\}$, we denote the marginal output probability by:
$p_{j}=\sum_{\mathbf{i}}p_{\mathbf{i}j}$. We define the joint probability
between the output state and the first input bit (the `WW bit') as,
$p_{i_{\rm ww},j}\equiv\sum_{i_{\rm wp}}p_{(i_{\rm ww},i_{\rm wp})j}$ and similar for
$p_{i_{2,}j}$. We define conditional probabilities such as $p_{i_{\rm ww}|j}$ and
$p_{i_{\rm ww}|j_{2}}$ as well. We use
\begin{equation}
\mathbf{r}_{\mathbf{i}}=\mathbf{r}_{i_{\rm ww},i_{\rm wp}}=\left( d_0, i_{\rm wp}d_{\rm wp},
i_{\rm ww}d_{\rm ww}\right)
\end{equation}
for the Bloch representation of the $\mathbf{i}$-th input state, and $\mathbf{R}_{j}= \left( x_j,y_j,z_j \right)$ for that of $A_j$ (the operator corresponding $j$-th measurement outcome), which is possible due to Lemma 1 above.
In this notation we have:
\begin{eqnarray}
\label{eq:prob_eqs}
p_{\mathbf{i}j} &= & \frac{\mu_{j}}{8}(1+d_{0}x_{j}+i_{\rm wp}y_{j}d_{\rm wp}+i_{\rm ww}z_{j}d_{\rm ww}), \\
p_{i_{\rm ww}j} & = & \frac{\mu_{j}}{4}(1+d_{0}x_{j}+i_{\rm ww}z_{j}d_{\rm ww}). 
\end{eqnarray}
Given that an output $j$ has occurred, the most probable value, $i_{\rm ww}$, of $b_{\rm ww}$ is that for which 
$p_{i_{\rm ww},j}>p_{-i_{\rm ww},j}$ The special case when these joint probabilities are equal needs to be treated slightly differently, but the results for the generic state can be shown to apply to it by continuity, and we shall not treat it here. From this follows that the total probability for correct WW inference is 
\be
P_{\rm WW}=\sum_{j}\max\left\{ p_{i_{\rm ww},j},p_{-i_{\rm ww},j}\right\}. 
\label{eq:PWW}
\ee
By Eq.
(\ref{eq:prob_eqs}),
\begin{equation}
\max\left\{
p_{i_{\rm ww},j},p_{-i_{\rm ww},j}\right\}
=\frac{\mu_{j}}{4}(1+d_{0}x_{j}+\left|z_{j}\right|d_{\rm ww}),
\end{equation}
therefore
\begin{equation}
P_{\rm WW}=\frac{1}{2}\left(1+\sum\frac{\mu_{j}}{2}\left|z_{j}\right|d_{\rm ww}\right),
\label{eq:12}
\end{equation}
where we have used $\frac{1}{2}\sum\mu_{j}=1,\sum x_{j}=0$. Similarly,
\begin{equation}
P_{\rm WP}=\frac{1}{2}\left(1+\sum\frac{\mu_{j}}{2}\left|y_{j}\right|d_{\rm wp}\right).
\label{eq:13}
\end{equation}
Combining these two equations, we have
\begin{equation}
\begin{split}
&\left(\frac{2P_{\rm WW}-1}{d_{\rm ww}}\right)^{2}+\left(\frac{2P_{\rm WP}-1}{d_{\rm wp}}\right)^{2}= \\
&\left(\sum\frac{\mu_{j}}{2}\left|z_{j}\right|\right)^{2}+\left(\sum\frac{\mu_{j}}{2}\left|y_{j}\right|\right)^{2} \leq \\
&\sum\frac{\mu_{j}}{2}\left(z_{j}^{2}+y_{j}^{2}\right) \leq \sum\frac{\mu_{j}}{2}=1.
\end{split}
\end{equation}
The last inequality follows from Jensen's inequality (for the concave function $f(x)=x^{2}$). $\blacksquare$

Let us now characterize the {\em Pareto-optimal POVMs}. 
The inequality becomes an equality iff $|z_{j}|=const.\equiv z_0,|y_{j}|=const.\equiv y_0,z_0^{2}+y_0^{2}=1$. 
Clearly, we are free to choose $y_0 \in [0,1]$, and let $z_0=\sqrt{1-y_0^2}$. Then the Bloch vectors $\mathbf{R}_j$ can
take the values $(0, \pm y_0, \pm z_0)$, with corresponding weights $\mu_{\pm,\pm}$.
The condition $\sum \mu_j \mathbf{R}_j = 0$ implies $\mu_{++}=\mu_{--},~\mu_{+-}=\mu_{-+}$
while $\sum \frac{1}{2}\mu_j = 1$ implies $\mu_{++}+\mu_{+-}=1$. Therefore, the optimal POVMs are represented by
rectangles residing on a great circle in the $y-z$ plane, with sides parallel to the rectangle of input states, and weights satisfying: 
\be 
\mu_{++}=\mu_{--}=\mu,~~\mu_{+-}=\mu_{-+}=1-\mu;~~\mu \in [0,1].
\label{eq:18}
\ee
For the optimal POVMs, Eqs. (\ref{eq:12}, \ref{eq:13}) simplify to 
\be
P_{\rm WW}=\frac{1}{2}\left(1+z_0 d_{\rm ww}\right),~~P_{\rm WP}=\frac{1}{2}\left(1+y_0 d_{\rm wp}\right).
\label{eq:opt}
\ee

\textbf{Bounds on $I_{\rm WW}+I_{\rm WP}$}

\begin{figure}[t]
\includegraphics[width=5cm]{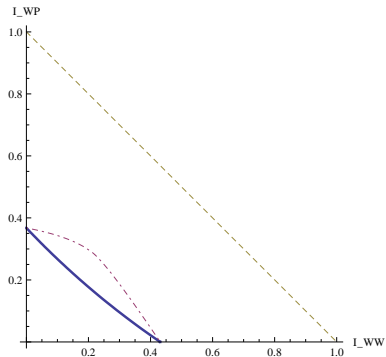}
\caption{Comparison of the exact tradeoff given by Eq. \eqref{eq:5} (solid line) with the
one given by Eq.\eqref{eq:1b} (dashed); the dot-dashed line shows $I_{\rm in-out}-I_{\rm WW}$:
the gap between it and the solid line is the contribution of $I_{\rm WW:WP}$
to $I_{in-out}$. All of these relations generalize naturally to
the three dimensional case ($I_{\rm WW}+I_{\rm WP}+I_{\rm WM}$).}
\label{fig:1b}
\end{figure}

We note that by Holevo's theorem\cite{nielsen521qca} $I_{\rm in-out}\leq S(\sum p_i \rho_i)-\sum p_i S(\rho_i) = H_2(\frac{1+d_0}{2})$, where $S$ denotes the Von-Neumann entropy, and $\rho_i~,p_i$ denote the $i$th initial state and its a priori probability, respectively. However, this bound is not tight in the present situation.
Together with Eq. \eqref{eq:26}, this implies the following bound: 

\begin{equation}
I_{\rm WW}+I_{\rm WP}\leq H_{2}(\frac{1+d_{0}}{2}),\label{eq:1b}
\end{equation}
where $d_{0}=\sqrt{1-\left(d_{\rm WW}^{2}+d_{\rm WP}^{2}\right)}.$ As mentioned there, this
bound is \emph{not} tight. However, as explained following Eq. \eqref{eq:10}, the
$P_{\rm WW}-P_{\rm WP}$ Pareto-frontier is also the frontier for $I_{\rm WW}-I_{\rm WP}$,
as the \emph{I}s are monotonic functions of the \emph{P}s. Thus, Eq. \eqref{eq:5}
implicity gives the exact tradeoff between the \emph{I}s. 

{\bf WW-WP-WM experiments}

The Von Neumann projective measurement depicted in Fig. \ref{fig:beamsplit_tunable} can roam on the entire ellipsoid, provided the input phase on the output BS be tunable, as well as its bias. Conversely, any output beam splitter is
Pareto-optimal. 
The WW detector generalizes to a two detector scheme, with detectors with efficiencies $E_{1},E_{2}$
(Fig. \ref{fig:multiple_bs}), yielding:
$\frac{P_{\rm WW}-\frac{1}{2}}{d_{\rm ww}}=E_{1},~ 
\frac{P_{\rm WP}-\frac{1}{2}}{d_{\rm wp}}=\sqrt{1-E_{1}^{2}}E_{2},~
\frac{P_{WM}-\frac{1}{2}}{d_{\rm wm}}=\sqrt{(1-E_{1}^{2})(1-E_{2}^{2})}$.

\end{document}